\documentclass[twocolumn,showpacs,preprintnumbers,superscriptaddress,amsmath,amssymb, prl]{revtex4-1}
\usepackage{graphicx}
\usepackage{dcolumn}
\usepackage{bm}

\begin{document}

\preprint{Ver. \today}

\title{Semimetallic "Electride Bands" Derived from Interlayer Electrons in Quasi-Two-Dimensional Electride Y$_2$C}

\author{Koji~Horiba}
\email{horiba@post.kek.jp}
\affiliation {Condensed Matter Research Center (CMRC) and Photon Factory (PF), Institute of Materials Structure Science, High Energy Accelerator Research Organization (KEK), Tsukuba 305-0801, Japan}

\author{Ryu~Yukawa}
\affiliation {Condensed Matter Research Center (CMRC) and Photon Factory (PF), Institute of Materials Structure Science, High Energy Accelerator Research Organization (KEK), Tsukuba 305-0801, Japan}

\author{Taichi~Mitsuhashi}
\affiliation {Condensed Matter Research Center (CMRC) and Photon Factory (PF), Institute of Materials Structure Science, High Energy Accelerator Research Organization (KEK), Tsukuba 305-0801, Japan}

\author{Miho~Kitamura}
\affiliation {Condensed Matter Research Center (CMRC) and Photon Factory (PF), Institute of Materials Structure Science, High Energy Accelerator Research Organization (KEK), Tsukuba 305-0801, Japan}

\author{Takeshi~Inoshita}
\affiliation {Materials Research Center for Element Strategy (MCES), Tokyo Institute of Technology, Yokohama 226-8503, Japan}

\author{Noriaki~Hamada}
\affiliation {Department of Physics, Tokyo University of Science, Noda 278-8510, Japan}

\author{Shigeki~Otani}
\affiliation {National Institute for Materials Science (NIMS), Tsukuba 305-0044, Japan}

\author{Naoki~Ohashi}
\affiliation {National Institute for Materials Science (NIMS), Tsukuba 305-0044, Japan}

\author{Sachiko~Maki}
\affiliation {Materials Research Center for Element Strategy (MCES), Tokyo Institute of Technology, Yokohama 226-8503, Japan}

\author{Jun-ichi~Yamaura}
\affiliation {Materials Research Center for Element Strategy (MCES), Tokyo Institute of Technology, Yokohama 226-8503, Japan}

\author{Hideo~Hosono}
\affiliation {Materials Research Center for Element Strategy (MCES), Tokyo Institute of Technology, Yokohama 226-8503, Japan}
\affiliation {Materials and Structures Laboratory, Tokyo Institute of Technology, Yokohama 226-8503, Japan}

\author{Youichi~Murakami}
\affiliation {Condensed Matter Research Center (CMRC) and Photon Factory (PF), Institute of Materials Structure Science, High Energy Accelerator Research Organization (KEK), Tsukuba 305-0801, Japan}

\author{Hiroshi~Kumigashira}
\affiliation {Condensed Matter Research Center (CMRC) and Photon Factory (PF), Institute of Materials Structure Science, High Energy Accelerator Research Organization (KEK), Tsukuba 305-0801, Japan}

\date{\today}

\begin{abstract}

Two-dimensional (2D) electrides are a new concept material in which anionic electrons are confined in the interlayer space between positively charged layers. We have performed angle-resolved photoemission spectroscopy measurements on Y$_2$C, which is a possible 2D electride, in order to verify the formation of 2D electride states in Y$_2$C. We clearly observe the existence of semimetallic "electride bands" near the Fermi level, as predicted by {\it ab initio} calculations, conclusively demonstrating that Y$_2$C is a quasi-2D electride with electride bands derived from interlayer anionic electrons.

\end{abstract}

\maketitle

Electrides are ionic crystals in which electrons serve as anions \cite{Dye1, Dye2}. They have recently attracted considerable attention as a new class of low-dimensional materials. Aside from generating scientific interest, electrides hold promise as novel engineering materials. The breakthrough discovery of the first room-temperature- and air-stable inorganic electride, [Ca$_{24}$Al$_{28}$O$_{64}$]$^{4+}$$\cdot$4e$^-$ (C12A7) \cite{C12A7} has opened up a new avenue in the application of electrides, capitalizing on their low work functions, including their use as electron injection layers in organic light emitting diodes \cite{OLED} and as catalysts for ammonia synthesis \cite{NH3_1, NH3_2}.

\begin{figure}
\begin{center}
\includegraphics[width=0.95\linewidth]{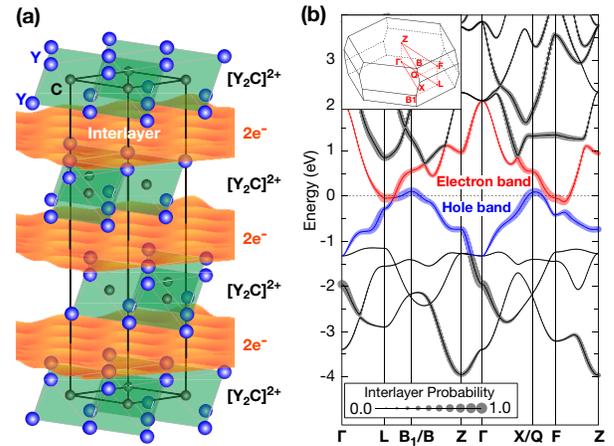}
\end{center}
\caption{(a) Crystal structure of Y$_2$C together with the schematic of the anionic electron layer. Solid thick lines represent the conventional hexagonal unit cell of rhombohedral Y$_2$C. {\it Ab initio} calculations predict that anionic electrons 2e$^-$ are confined in the interlayer space between the [Y$_2$C]$^{2+}$ sheets. (b) Band structure of Y$_2$C calculated by the PAW method. The interlayer probability, i.e., the probability of the electron being found in the interlayer space, is represented by the size of the filled circles. Semimetallic "electride bands" are formed near $E_F$. The inset shows the BZ for the rhombohedral lattice of Y$_2$C together with the symmetry points.}
\label{figure1}
\end{figure}

The electronic properties of electrides are expected to depend critically on the topology and dimensionality of the void space confining the anionic electrons. For many years after their discovery, all the known electrides were found to be either quasi-zero-dimensional (electrons confined in cages) \cite{C12A7} or quasi-one-dimensional (electrons confined in filamentary channels) \cite{1D, Quasi}. A recent study has reported the synthesis of the first possible two-dimensional (2D) electride, Ca$_2$N \cite{Ca2N}, and further, the possible existence of 2D electride states in Ca$_2$N has been supported by {\it ab initio} calculations \cite{Ca2N, Walsh_and_Scanlon}. Subsequently, Inoshita {\it et al}. have predicted Y$_2$C as another possible candidate for 2D electrides \cite{Inoshita_PRX}. These materials have a unique layered structure, as shown schematically in Fig.~\ref{figure1}(a): One (two) anionic electron(s) per Ca$_2$N (Y$_2$C) unit cell ([Ca$_2$N]$^+$$\cdot$e$^-$ or [Y$_2$C]$^{2+}$$\cdot$2e$^-$) is (are) confined in the interlayer space between positively charged [Ca$_2$N]$^+$ ([Y$_2$C]$^{2+}$) sheets. {\it Ab initio} calculations have predicted that the confined electrons freely move inside this 2D void space and consequently form "electride bands" near the Fermi level ($E_F$) \cite{Ca2N, Walsh_and_Scanlon, Inoshita_PRX, Tada_Cal}. In other words, the 2D confinement of anionic electrons can be regarded as an ideal quantum-well state of free electrons, which provides an excellent research platform for studying the quantum physics of 2D-electron systems \cite{QHE, LAOSTO} as well as possible applications for future electronic devices \cite{HEMT}.

The crucial issue in the research of 2D electrides is whether or not the material concept is actually realized in Ca$_2$N and/or Y$_2$C. In this context, Lee {\it et al}. have reported detailed transport measurements of single-crystalline Ca$_2$N \cite{Ca2N}. The realization of the material concept of 2D electrides in Ca$_2$N is supported by two unique behaviors: very high electron mobility even at very high electron concentrations ($\sim$10$^{22}$~cm$^{-3}$) corresponding to that for the assumed formula [Ca$_2$N]$^+$$\cdot$e$^-$, and negative magnetic resistivity upon magnetic-field application along the normal to the confined space. These experimental results are fully consistent with {\it ab initio} calculations. Furthermore, subsequent studies have also revealed that Ca$_2$N is a nearly-2D electride \cite{Ca2N, Walsh_and_Scanlon, Inoshita_PRX, Tada_Cal, Fang_Cal_Ca2N}.

On the other hand, as regards the possible 2D electride Y$_2$C, transport, x-ray photoemission spectroscopy, and magnetic susceptibility measurements have been conducted only for polycrystalline samples of Y$_2$C \cite{Y2C}. However, such experiments can at best be considered as indirect probes of electronic structures, although these results appear to be consistent with {\it ab initio} calculations. In order to credibly prove that this material is indeed a 2D electride (in which anionic electrons are confined two-dimensionally in the interlayer space and consequently "electride bands" are formed near $E_F$), it is necessary to directly probe the electronic band structure.

In this Letter, we provide experimental evidence that Y$_2$C is indeed a quasi-2D electride having semimetallic band structures by using angle-resolved photoemission spectroscopy (ARPES). As shown in Fig.~\ref{figure1}, the existence of mobile anionic electrons in the interlayer space has a one-to-one correspondence with the formation of "electride bands" near $E_F$ as per band-structure calculations. That is, the direct observation of these electride bands near $E_F$ is the direct proof of the realization of the (quasi-) 2D electride states in Y$_2$C. In the study, we clearly observed that electron and hole electride bands exist near $E_F$, and that these semimetallic bands form Fermi surfaces (FSs), as predicted by {\it ab initio} calculations \cite{Inoshita_PRX, Y2C}. Furthermore, the observed band structure is in good agreement with the results of {\it ab initio} calculations. This agreement between the theoretical and experimental results clearly demonstrates that Y$_2$C is a quasi-2D electride having semimetallic electride bands crossing $E_F$.

Single crystals of Y$_2$C were grown by means of the floating zone method under an argon pressure of 0.4~MPa. The details of the growth conditions and characterizations are described elsewhere \cite{Growth_Y2C}. Since Y$_2$C is highly reactive with ambient oxygen and water vapor, the samples were affixed to the ARPES holder with a silver conductive epoxy adhesive in a purified Ar-filled glove box and then moved to an ultrahigh vacuum (UHV) chamber for ARPES measurements with the use of a specially designed transfer vessel.

\begin{figure}
\begin{center}
\includegraphics[width=0.95\linewidth]{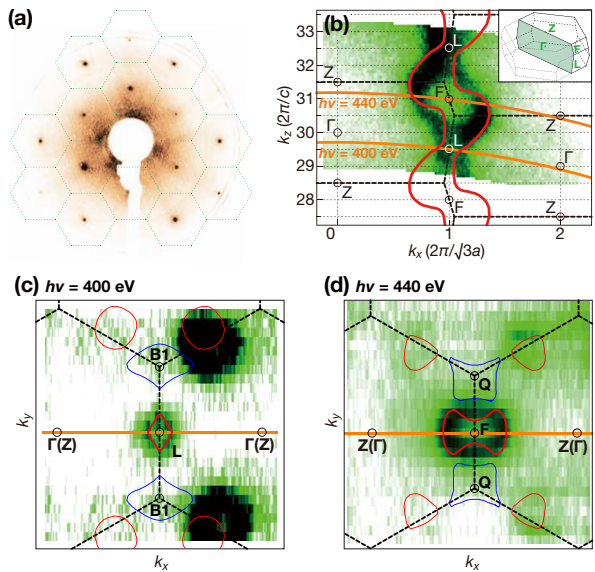}
\end{center}
\caption{(a) LEED pattern for a cleaved surface of Y$_2$C single crystal acquired at an electron energy of 124~eV. Dotted lines represent the hexagonal surface BZ. (b) Out-of-plane FS map in the ${\Gamma}ZFL$ plane obtained by varying the excitation photon energy from 380~eV to 500~eV. The FS map was obtained by plotting the ARPES intensity within the energy window of $\pm$100~meV from $E_F$. The dotted lines indicate the BZ boundaries, while red solid lines indicate the electron FS obtained by FLAPW calculations. The orange lines represent the $\bm{k}$-path passing through the $L$ point (at a photon energy of 400~eV) and the $F$ point (at a photon energy of 440~eV). (c), (d) In-plane FS maps acquired at constant photon energies of 400~eV (c) and 440~eV (d) by changing emission angles. Calculated FSs at in-plane ($k_x$--$k_y$ plane) crossing the $L$ point ($k_z$~=~29.5~[2$\pi/c$]) and the $F$ point ($k_z$~=~31.0~[2$\pi/c$]) are superimposed on (c) and (d), respectively. Red and blue lines show electron and hole pockets, respectively.}
\label{figure2}
\end{figure}

ARPES measurements were carried out using horizontally polarized synchrotron radiation light at the BL-2A MUSASHI beam line of the Photon Factory, KEK. The samples were cleaved {\it in situ} under a UHV of $\sim$10$^{-8}$~Pa at a low temperature ($\sim$20~K). As shown in Fig.~\ref{figure2}(a), the low-energy electron diffraction (LEED) pattern exhibits a hexagonal arrangement of spots corresponding to the reciprocal lattice of the $R\overline{3}m$ space group \cite{Y2Ccrys_1, Y2Ccrys_2}, thus indicating the obtainment of a clean and well-ordered Y$_2$C surface. For the present ARPES study, soft x rays (SX) with a large probing depth were used to probe the intrinsic electride states because the anionic electron layer is expected to lie buried below the top [Y$_2$C]$^{2+}$ layer. The sample temperature during the ARPES measurements and the total energy resolution were set to 20~K and 200~meV at a photon energy ($h\nu$) of around 400~eV, respectively.

We carried out density-functional electronic structure calculations of Y$_2$C with the generalized gradient approximation (GGA) of the exchange-correlation functionals. Two different methods were employed: (1) the full-potential linearized augmented plane wave (FLAPW) method, as implemented in the all-electron band structure calculation package (ABCAP) \cite{PRB14} and (2) the projector augmented wave (PAW) method using a plane wave basis set, as implemented in the Vienna {\it ab initio} simulation package (VASP) \cite{PRB15, PRB16, PRB17, PRB18}. We used the Perdew-Burke-Ernzerhof (PBE) \cite{PRB20} and Perdew and Wang (PW91) \cite{PRB21} GGA potentials for (1) and (2), respectively. In the FLAPW calculations, the cutoff energies were set to be 163~eV and 653~eV for the wave functions and charge/potential, respectively, and momentum ($\bm{k}$-) integration was performed in the conventional (hexagonal) Brillouin zone (BZ) using a $12\times12\times4$ $\Gamma$-centered mesh. For the PAW calculations, the plane waves were cut off at 800 eV, and integration over the primitive (rhombohedral) BZ was performed with a $17\times17\times17$ $\Gamma$-centered mesh. The structure was relaxed using the PAW method until the force acting on each ion became less than 0.7~meV/\AA. The optimized structure was used for FLAPW calculations. The band structures obtained by the two methods were in excellent agreement with each other.

Before discussing the ARPES results, we explain the vital difference in the electronic structures of the two types of electrides, nitrides (Ca$_2$N) and carbides (Y$_2$C) in terms of the band-structure calculation. Since the oxidation numbers of Ca and N are +2 and $-$3, respectively, Ca$_2$N has one excess electron per unit cell [Ca$_2$N]$^+$$\cdot$e$^-$, which is two-dimensionally confined in the interlayer space and forms a half-filled interlayer (electride) band. As a result of the nearly 2D nature of the anionic electrons confined in the interlayer space, the FS, which is mainly derived from the {\it metallic} electride band near $E_F$, is located at the BZ center and is highly cylindrical. That is, Ca$_2$N is a nearly-2D electride metal \cite{Ca2N, Walsh_and_Scanlon, Inoshita_PRX, Tada_Cal}.

While the oxidation numbers of Y and C (+3 and $-$4, respectively) indicate that Y$_2$C has two excess electrons per unit cell [Y$_2$C]$^{2+}$$\cdot$2e$^-$ and is therefore a band insulator, our {\it ab initio} calculations reveal that Y$_2$C is a semimetal: An almost fully occupied hole-like electride band and a slightly occupied electron-like electride band overlap each other near $E_F$, as shown in Fig.~\ref{figure1}(b). Here, the size of the circles in Fig.~\ref{figure1}(b) indicates the interlayer probability of the calculated eigenstates, i.e., the probability that the electron is found inside a sphere of radius 1.7~{\AA} at the interstitial site between Y$_2$C layers. Note the strong interlayer (electride) character of the two semimetallic bands near $E_F$. The FSs for the two bands are located near the BZ boundary and are shaped as rugged cylinders (not shown), thereby reflecting the rather three-dimensional (3D) nature of the anionic electrons of Y$_2$C over those of Ca$_2$N \cite{Ca2N, Walsh_and_Scanlon, Inoshita_PRX, Tada_Cal, Y2C, Inoshita_PRB}. Thus, according to the band-structure calculation, Y$_2$C is a quasi-2D electride having semimetallic band structures.

\begin{figure}
\begin{center}
\includegraphics[width=0.95\linewidth]{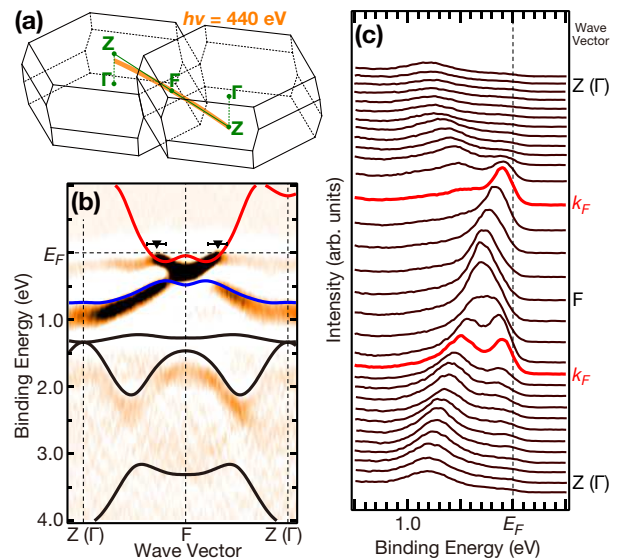}
\end{center}
\caption{(a) Measured $\bm{k}$-path passing through the $F$ point in the rhombohedral BZ for which ARPES measurements were carried out at $h\nu$~=~440~eV, which corresponds to the orange line in Fig.~\ref{figure2}(d). (b) Experimental band structure obtained by plotting the second derivative of the ARPES spectra acquired along the $\bm{k}$-path in (a). The energy bands along the $Z$--$F$--$Z$ direction calculated by the FLAPW method are superimposed by solid lines. Filled triangles indicate the Fermi momentum ($k_F$) determined by ARPES spectra \cite{Suppl}. (c) ARPES spectra near $E_F$ and around the $F$ point. Thick red lines indicate the ARPES spectra at $k_F$.}
\label{figure3}
\end{figure}

In order to directly observe the electride band near $E_F$ as evidence of the realization of quasi-2D electride states in Y$_2$C, we performed ARPES measurements. Figure~\ref{figure2}(b) shows an experimental out-of-plane FS obtained by plotting the ARPES intensity at $E_F$ in the ${\Gamma}ZFL$ plane for varying excitation photon energies. A meandering FS derived from the electron band is clearly observed at the BZ boundary, following the periodicity of the rhombohedral BZ. As can be observed in Fig.~\ref{figure2}(b), the overall shape of the observed FS is in excellent agreement with the results of our calculation. The observed strong modulation in the FS along the momentum perpendicular to the surface ($k_z$) strongly suggests the increased 3D character of the electride states in Y$_2$C in comparison with those in Ca$_2$N \cite{Ca2N}.

The observation of $k_z$-modulation in the FS indicates that the present SX-ARPES clearly reveals a bulk electronic structure in Y$_2$C. This means that the large probing depth of the SX-ARPES enables us to examine the buried anionic electrons in the positively charged layer [see Fig.~\ref{figure1}(a)]. Therefore, we next measured the in-plane FS at a constant photon energy by varying the emission angle; the corresponding results are shown in Figs.~\ref{figure2}(c) and (d). The FS mapping acquired at $h\nu$~=~400~eV and 440~eV traces the spherical surface of the $\bm{k}$-space through the high-symmetry $L$ and $F$ points [each profile in the ${\Gamma}ZFL$ plane is shown as an orange line in Fig.~\ref{figure2}(b)], respectively. Therefore, these FS maps could be exactly compared with the band-structure calculation only around the $L$ and $F$ points because the FS maps shown in Figs.~\ref{figure2}(c) and (d) represent the projections of the spherical surface on the corresponding $k_x$--$k_y$ plane. The observed electron FSs around the $L$ and $F$ points are in good agreement with the corresponding prediction of the theoretical calculations. On the other hand, the topology of the hole FSs that are predicted to be located around the $B1$ and $Q$ points are not clearly resolved in the FS maps, probably because of the matrix element effect, although the hole band crossing $E_F$ itself is observed in the ARPES spectra \cite{Suppl}.

\begin{figure}
\begin{center}
\includegraphics[width=0.95\linewidth]{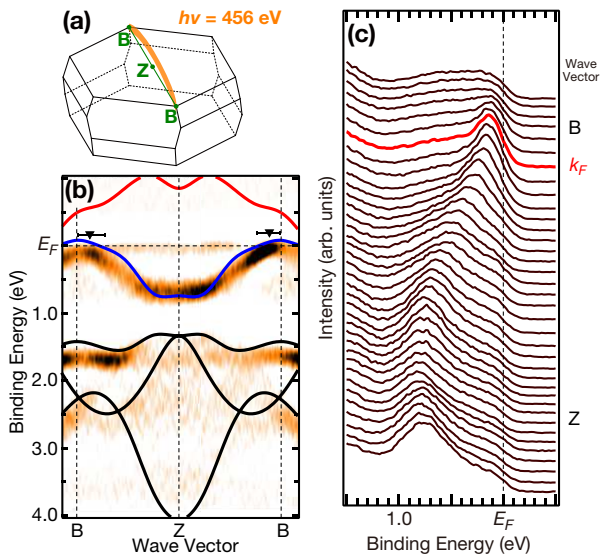}
\end{center}
\caption{(a) Measured $\bm{k}$-path passing through the $B$ point of the rhombohedral BZ for which ARPES measurements were carried out at $h\nu$~=~456~eV. (b) Experimental band structures obtained by plotting the second derivative of ARPES spectra. Solid lines indicate the energy bands calculated by the FLAPW method along the $B$--$Z$--$B$ direction. Filled triangles indicate the $k_F$ points determined by ARPES spectra \cite{Suppl}. (c) ARPES spectra near $E_F$ and around the $B$ point. The thick red line indicates the ARPES spectrum at $k_F$.}
\label{figure4}
\end{figure}

For investigating the electronic structures and their correspondence with the theoretical calculations in detail, we measured the energy-band dispersion along the representative high-symmetry lines. Figures~\ref{figure3}(b) and \ref{figure4}(b) show the band dispersions determined by ARPES along the $Z$--$F$--$Z$ and $B$--$Z$--$B$ directions [along the orange line in the BZ shown in Figs.~\ref{figure3}(a) and \ref{figure4}(a)], respectively. It is clear that electron and hole electride bands exist near $E_F$, leading to the formation of a small electron- and hole-pocket around the $F$ and $B$ points, which pocket is more clearly seen in the ARPES spectra shown in Figs.~\ref{figure3}(c) and \ref{figure4}(c), respectively. According to calculations, the highly dispersive bands around 1.5--2.5~eV mainly consist of the C~2$p$ orbital, while the dispersive bands in the energy range from near-$E_F$ to around 1~eV are mainly derived from the anionic electrons in the interlayer space. As a result of the considerable hybridization between the electride band near $E_F$ and Y~4$d$ states above $E_F$, semimetallic electride bands are formed near $E_F$. Although there is a quantitative discrepancy between the experiment and calculation around the $F$ points (which we discuss in detail later), the overall band structures observed by ARPES are in good agreement with theoretical band structures. Consequently, these results indicate that Y$_2$C is most evidently a quasi-2D electride having semimetallic electride bands derived from anionic electrons confined in the interlayer space.

Next, we discuss the possible influence of certain surface states characteristic of electrides. A recent study has reported the observation of metallic band structures in Ca$_2$N via ARPES using surface-sensitive vacuum ultraviolet light \cite{ARPES_Ca2N}. In this study, the size of the FS determined by ARPES is considerably smaller than that predicted theoretically. Based on their observations, the authors have argued that the reduction in the size of the FS originates from the shift of the chemical potential due to the depletion of anionic electrons on the cleaved surface. In contrast, in the present study, the size of the FS estimated from ARPES, as well as the width of the electride band, are in good agreement with the theoretical predictions, strongly suggesting that the present ARPES using bulk-sensitive SX is nearly free from the influence of the surface states characteristic of electrides. These results suggest that ARPES with a large probing depth is necessary to reveal the intrinsic bulk electronic structures of electrides.

Finally, we discuss the observed quantitative discrepancy in the electride band of Y$_2$C between the experiment and calculation. The good agreement of the bandwidth between the two strongly suggests that the electron-electron correlation among interlayer electrons is as negligibly weak in Y$_2$C as in Ca$_2$N \cite{ARPES_Ca2N}. Although the overall band structure is well described by the GGA calculation, a slight quantitative discrepancy is observed at the $F$ point: In the experiment, the bottom of the electron band appears to touch the top of the hole band, while the two bands overlap in the calculations. This discrepancy may arise from a somewhat inaccurate positioning of the Y~4$d$ bands by GGA. In fact, GGA+$U$ and hybrid functional (HSE06) calculations (not shown) predict an upshift of the Y~4$d$ bands when compared with that for GGA, reducing the band overlap at the $F$ point and bringing the theoretical predictions closer to our experimental results.

The semimetallicity in Y$_2$C, together with hybridization, albeit weak, with the Y~4$d$ bands, results in a rather heavy in-layer effective mass in the vicinity of $E_F$ and a strong peak in the density of states at $E_F$ \cite{Inoshita_PRB}. Therefore, this non-trivial band structure may give rise to interesting physical properties. For example, it is predicted that Y$_2$C is close to a Stoner-type ferromagnetic instability, which is induced by interlayer electrons rather than the $d$ electrons of Y \cite{Inoshita_PRB}. In fact, the actual value of the magnetic susceptibility is still in question \cite{Y2C, Growth_Y2C} because there is strong sample-to-sample variation, which can be attributed to the complex phase diagram of the Y--C system. Considering the unique semimetallic band structure, there is a possibility that the intriguing physical properties of Y$_2$C originate from the delicate balance between the electron and hole electride bands near $E_F$. Thus, further investigations of the detailed electronic structures and their relation to the physical properties of Y$_2$C, as well as other 2D electrides, are required.

In summary, we performed ARPES measurements on Y$_2$C to test whether or not it is a 2D electride. We clearly observed semimetallic "electride bands" near $E_F$ via ARPES, as predicted from {\it ab initio} theoretical calculations. Furthermore, these band dispersions were in good agreement with the results of the calculations. The good agreement between theory and experiment proves that Y$_2$C is a quasi-2D electride having a semimetallic band structure.

\acknowledgements
The authors are very grateful to Yoshitake Toda, Satoru Matsuishi, and Tomofumi Tada for fruitful discussions. This work was supported by the MEXT Elements Strategy Initiative to Form Core Research Center as well as Grants-in-Aid for Scientific Research (Nos. B25287095, 16H02115, and 16K05033) from the Japan Society for the Promotion of Science (JSPS). M.K. acknowledges financial support from JSPS for Young Scientists. H.H. acknowledges the support of the JST ACCEL Program. The work at KEK-PF was performed under the approval of the Program Advisory Committee (Proposal Nos. 2013S2-002 and 2015S2-005) at the Institute of Materials Structure Science, KEK.


\begin{references}
\bibitem{Quasi} We use the prefix 'quasi' to emphasize that we are referring to bulk low-dimensional materials, with anisotropic electronic structures, rather than molecules, atomic chains, or atomically thin slabs such as graphene.
\end{references}
\end{document}